\documentclass[nofootinbib,amsmath,amssymb,aps,twocolumn,prd,preprintnumbers]{revtex4}
\usepackage{graphicx} 
\usepackage{dcolumn} 
\usepackage{bm} 
\usepackage{xcolor} 
\usepackage[colorlinks=true,linktoc=all]{hyperref} 
\usepackage{mathtools}

\begin{document}
\preprint{IPARCOS-UCM-23-066}

\title{Cosmology in gravity models with broken diffeomorphisms}

\author{Antonio G. Bello-Morales}%
 \email{antgon12@ucm.es}
\affiliation{Departamento de F\'{\i}sica Te\'orica and\\ Instituto de F\'{\i}sica de Part\'{\i}culas y del Cosmos (IPARCOS-UCM), Universidad Complutense de Madrid, 28040 
Madrid, Spain}
\author{Antonio L. Maroto}%
 \email{maroto@ucm.es}
\affiliation{Departamento de F\'{\i}sica Te\'orica and\\ Instituto de F\'{\i}sica de Part\'{\i}culas y del Cosmos (IPARCOS-UCM), Universidad Complutense de Madrid, 28040 
Madrid, Spain}

\date{\today}

\begin{abstract}
We study the cosmological implications of gravity models which break diffeomorphisms (Diff) invariance down to transverse diffeomorphisms (TDiff). We start from the most general gravitational action involving up to quadratic terms in  derivatives of the metric tensor and identify TDiff models as the only stable theories consistent with local gravity tests. These models propagate an additional scalar graviton and although they are indistinguishable from GR at the post-Newtonian level, their cosmological dynamics exhibits a rich phenomenology. Thus we show that the model includes standard $\Lambda$CDM as a solution when the extra scalar mode is not excited, but different cosmological evolutions driven by the new term are possible. In particular, we show that for a soft Diff breaking, the new contribution always behaves as a cosmological constant at late times. When the extra contribution is not negligible, generically its 
evolution either behaves as dark energy or tracks the dominant background component. Depending on the initial conditions, solutions in which the universe evolves from an expanding to a contracting phase, eventually recollapsing are also possible.

\end{abstract}

\maketitle

\section{Introduction}
The principle of General Covariance \cite{Weinberg:1972kfs,Gaul:1999ys}, i.e. "the laws of physics retain the same form under arbitrary coordinate transformations" is one of the cornerstones of the theory of General Relativity (GR). This principle, on one hand,  governs the interactions with the gravitational field, selecting the allowed couplings to matter and, on the other, sets the dynamics of the gravitational field itself. Thus,  very much as for local gauge  symmetries, invariance under diffeomorphisms (Diff) allows to eliminate from the physical spectrum of the (linearized) theory all the degrees of freedom
contained in the metric tensor except for the massless spin-2 graviton.

Despite the fundamental nature of this principle, in recent years a lot of activity has been taking place on the possibility of building 
consistent theories of gravity which break Diff invariance.
This has been  motivated in part by the success of unimodular gravity
\cite{Einstein, Unruh:1988in, Alvarez:2005iy, Carballo-Rubio:2022ofy} as a possible solution to the vacuum energy problem \cite{Ellis:2010uc,Jirousek:2023gzr}. Unimodular  gravity restricts  the determinant of the metric tensor to be a non-dynamical field thus  breaking Diff invariance  down to transverse diffeomorphisms (TDiff) \cite{Alvarez:2006uu, Lopez-Villarejo:2010uib}. As a matter of fact, it has been shown that it is TDiff invariance, rather than full Diffs,   the minimal symmetry required by unitarity in theories with a  massless spin-2 field \cite{vanderBij:1981ym}. Unimodular gravity is thus seen to propagate the same degrees of freedom as General Relativity and its field equations of motion are just Einstein equations supplemented with a cosmological constant term which appears as an integration constant \cite{Henneaux:1989zc,Kuchar:1991xd}.

Given the fundamental role of TDiff symmetry for the consistency of gravity theories, TDiff  models beyond unimodular gravity have also been explored  in which the metric determinant is a dynamical field  \cite{Alvarez:2006uu,Pirogov:2005im,Pirogov:2009hr,Pirogov:2011iq,Pirogov:2014lda,Barcelo:2017tes}. The spectrum of these theories includes a scalar graviton in addition to the standard
massless spin-2 graviton and some phenomenological implications have  been explored in  \cite{Pirogov:2011iq,Pirogov:2014lda}. 

Apart from these particular examples,  a more general effective field theory approach has been considered in \cite{Anber:2009qp}. There, the most general Lorentz invariant action up to quadratic terms in metric derivatives is obtained and the corresponding post-Newtonian (PPN) parameters   \cite{Will:2014kxa} are  explicitly worked out in some particular cases. The general conclusion suggests that  violations of Diff invariance are severly constrained by local
gravity experiments. However, certain combinations of terms could still be viable. As a matter of fact, models different from General Relativity are identified which nevertheless provide the same equations of motion in the weak field approximation. 

The breaking of Diff invariance in the couplings to matter have also been analyzed in  \cite{Alvarez:2009ga,Jirousek:2020vhy, Maroto:2023toq, Jaramillo-Garrido:2023cor}. Thus in \cite{Alvarez:2009ga} TDiff invariant models for spin-0 fields were studied  and potential violations of the weak equivalence principle (WEP) were identified. However, in \cite{Maroto:2023toq}, it was shown that in the geometric optics approximation it is possible to find models in which the three types of masses (inertial, active and passive) agree with those of standard Diff invariant theories thus evading the mentioned conflicts.

In this work we will focus on Diff breaking in cosmological contexts. We will start by identifying  TDiff invariant models which are compatible with Newtonian gravity in the weak field approximation and in addition 
have the same PPN parameters as GR. Even though the models under consideration are indistinguishable from GR in local gravity experiments, their non-linear dynamics can be very different. In particular, we will show that the presence of the extra gravitational degree of freedom  generates a wide range of new cosmological solutions.   

The paper is organized as follows: in Section II we consider the most general gravitational action up to terms with two metric derivatives and identify the 
consistent models. In Section III, we obtain the modified Einstein equations.
In Section IV we apply these results to Robertson-Walker backgrounds and show that the modified Friedmann equations can be rewritten as ordinary Friedmann equations with an additional effective perfect fluid contribution. Section V is devoted to the derivation of explicit solutions and in  Section VI we obtain a useful set of equations involving the effective equation of state of the new contribution. In Section VII, solutions are obtained in the subdominant regime in which the new effective energy contribution is negligible compared to that of standard matter and radiation. Section VIII is devoted to the opposite regimen in which the extra contribution is dominant.
In Section IX, we consider the general solution in which all the energy contributions are taken into account. In Section X, we study the stability of solutions and in Section XI we present the main 
conclusions of the work.

\section{Gravity with broken diffeomorphisms}
Following \cite{Anber:2009qp}, let us consider the most general expression for a global Lorentz invariant action for gravity in the metric formalism involving terms up to quadratic order in metric derivatives. 
\begin{align}
S_G=-\frac{1}{16\pi G}\int d^4x \left(\sum_{i=1}^5 f_i(g) \mathcal{L}_i + f_\Lambda(g)\right)
\label{SG}
\end{align}
where
\begin{align}
\begin{split}
    &\mathcal{L}_1 = -g^{\mu \nu} \Gamma^{\alpha}_{\mu \lambda} \Gamma^\lambda_{\nu \alpha}, \quad \mathcal{L}_3 = -g^{\mu \nu}g^{\rho \sigma} g_{\lambda \omega} \Gamma^\lambda _{\mu \rho} \Gamma^{\omega}_{\nu \sigma} \\
    &\mathcal{L}_2 = -g^{\mu \nu} \Gamma^{\alpha}_{\mu \nu} \Gamma^\lambda_{\lambda \alpha}, \quad  \mathcal{L}_4 = -g^{\mu \nu}g^{\rho \sigma} g_{\lambda \omega} \Gamma^\lambda _{\mu \nu} \Gamma^{\omega}_{\rho \sigma} \\ &\mathcal{L}_5 = -g^{\alpha \beta} \Gamma^{\lambda}_{\lambda \alpha} \Gamma^{\mu}_{\mu \beta} \label{five}
\end{split}
\end{align}
with $\Gamma^{\alpha}_{\mu \nu}$ the Christoffel symbols and $f_i(g)$ arbitrary functions \footnote{Notice that for $f_i(g)\propto \sqrt{g}, \forall i$, the action is invariant not only under global Lorentz transformations but also under global $GL(4, \mathbb{R})$ transformations} of the metric determinant $g=\vert \det g_{\mu\nu}\vert$.

Notice that the Einstein-Hilbert action \footnote{We are using  $(+,-,-,-)$ for the metric signature and the following definition for the Riemann tensor $R^{\rho}_{\;\;\sigma\mu\nu}=\partial_\mu\Gamma^\rho_{\nu\sigma}-\partial_\nu\Gamma^\rho_{\mu\sigma}+\Gamma^\rho_{\mu\lambda}\Gamma^\lambda_{\nu\sigma}-\Gamma^\rho_{\nu\lambda}\Gamma^\lambda_{\mu\sigma}$}
\begin{align}
S_{EH}=-\frac{1}{16\pi G}\int d^4x \sqrt{g}\, R   
\end{align}
is a particular case of the above general action, since it can be written up to total derivative terms as
\begin{align}
S_{EH}=-\frac{1}{16\pi G}\int d^4x \sqrt{g} \left(\mathcal{L}_2-\mathcal{L}_1\right) 
\end{align}

In the weak field approximation around the Minkowski background
\begin{equation}
    g_{\mu \nu} = \eta_{\mu \nu} + h_{\mu \nu} ; \quad |h_{\mu \nu}| \ll 1
\end{equation}
the linearized Einstein equations obtained from \eqref{SG} read \cite{Anber:2009qp}
\begin{align}
\begin{split}
    &(a_1-3a_3)\Box h^{\alpha \beta} + (-a_1+a_3-2a_4) (\partial^\alpha \partial_\gamma h^{\beta \gamma} + \partial^\beta \partial_\gamma h^{\alpha \gamma}) \\ + &(-a_2 +  2a_4)\eta^{\alpha \beta}\partial_\mu \partial_\nu h^{\mu \nu} + (-a_2 + 2a_4)\partial^\alpha \partial^\beta h \\ &+(a_2 - a_4 - a_5) \eta^{\alpha \beta} \Box h = 16\pi G T^{\alpha \beta}
\end{split}
\end{align}
where $h=h^\alpha_{\;\; \alpha}$, we have defined $a_i=f_i(g=1)$ and taken $a_\Lambda=0$. Assuming that diffeomorphisms invariance is only broken in the gravitational sector we can impose the energy-momentum tensor conservation $\partial_\alpha T^{\alpha \beta}=0$ so that we end up with
\begin{align}
    &2(a_3+a_4) \Box \partial_\beta h^{\alpha \beta} +(a_1+a_2-a_3)\partial^\alpha \partial_\mu\partial_\nu h^{\mu\nu} \nonumber \\
    &+ (a_5 - a_4) \partial^\alpha \Box h = 0 
\end{align}
Thus, as expected, in the particular case with
$a_1=-a_2=-1$ and $a_3=a_4=a_5=0$ we recover the standard linearized Einstein equations. However as shown in \cite{Anber:2009qp}, there are two additional special cases in which we can recover the linearized Einstein equation in a particular gauge i.e. $\Box \Bar{h}^{\alpha \beta} = 16\pi G T^{\alpha \beta}$. 
\begin{itemize}
    \item  $a_1=-a_2=-1$ and $a_3=a_5=0$ but $a_4\neq 0$, by using the trace reversed tensor $h^{\alpha \beta} = \Bar{h}^{\alpha \beta} - \eta^{\alpha \beta}\Bar{h}/2$
    \item  $a_1=-a_2=-1$ and $a_3=a_4=0$ but $a_5\neq 0$ by using the tensor $h^{\alpha \beta} = \Bar{h}^{\alpha \beta} - \eta^{\alpha \beta}\Bar{h}/4$.
\end{itemize}
Given the fact that these two models do not lead to physical consequences beyond GR at the linear level, they are, a priori, 
good candidates for a viable theory and we will concentrate on them in the following. 

At the quadratic order, the five terms in  \eqref{five} are not independent and can be written in terms of only four terms \cite{Alvarez:2006uu}
\begin{align}
 S=\int d^4x \mathcal{L}
\end{align}
where we have absorbed a $(16\pi G)^{-1/2}$ factor in a redefinition of the $h_{\mu\nu}$ field
that now becomes dimensionful, so that 
\begin{equation}
    \mathcal{L} = \mathcal{L}^I + \beta \mathcal{L}^{II} + a\mathcal{L}^{III} + b\mathcal{L}^{IV}
\end{equation}
with
\begin{align}
\begin{split}
    &\mathcal{L}^I = \frac{1}{4}\partial_\mu h^{\nu \rho} \partial^\mu h_{\nu \rho}  \quad \mathcal{L}^{II} = -\frac{1}{2}\partial_\mu h^{\mu \rho}\partial^\nu h^\nu_\rho  \\ \quad &\mathcal{L}^{III} = \frac{1}{2}\partial^\mu h \partial^\rho h_{\mu \rho}  \quad \mathcal{L}^{IV} = -\frac{1}{4}\partial_\mu h \partial^\mu h 
\end{split}
\end{align}

The standard Diff invariant Einstein-Hilbert action corresponds to $a=b=\beta=1$. Terms $\mathcal{L}^{III}$ and $\mathcal{L}^{IV}$ as well as the combination $\mathcal{L}^I + \mathcal{L}^{II}$ i.e. models with $\beta=1$ are invariant under transverse diffeomorphisms (TDiff) given infinitesimally by transformations  
\begin{align}
   \hat h_{\alpha\beta}(x)=h_{\alpha\beta}(x)-\xi_{\alpha , \beta}(x)-\xi_{\beta , \alpha}(x) 
\end{align}
such that $\partial_\alpha \xi^\alpha(x)=0$. 

It has been found that breaking TDiff symmetry by taking $\beta \neq 1$ introduces vector instabilities \cite{Alvarez:2006uu} in the solutions.

Thus, limiting ourselves to the two special cases mentioned above which reproduce GR in the weak field limit, the quadratic Lagrangian 
can be written as
\begin{equation}
   \mathcal{L} = \mathcal{L}^{I} + (1-2a_4)(\mathcal{L}^{II} + \mathcal{L}^{III}) + (1-a_4-a_5)\mathcal{L}^{IV} \label{modelterms}
\end{equation}
where we identify $\beta = a = 1 - 2a_4$ , $b = 1 - a_4 - a_5$. Thus, we see that in the $a_4\neq 0$ case TDiff invariance  is broken and we have instabilities as mentioned above. However the $a_4=0$, $a_5\neq 0$ case is TDiff invariant.  This theory propagates a scalar mode in addition to the two standard tensor modes of GR\footnote{For TDiff models it can be seen \cite{Alvarez:2006uu} that vector modes are not dynamical}. The corresponding Lagrangian for the scalar mode
reads \cite{Alvarez:2006uu}
\begin{align}
    \mathcal{L}_{S} = -\frac{\Delta b}{4}(\partial_\mu h)^2 = \frac{a_5}{4}(\partial_\mu h)^2 \label{int}
\end{align}
with 
\begin{align}
\Delta b=b-\frac{1-2a+3a^2}{2}=-a_5
\end{align}
Thus we must take $a_5 > 0$ in order to avoid ghost instabilities.

Regarding the coupling to matter, it has been shown that the most general TDiff invariant coupling to matter for the linearized theory takes the form \cite{Alvarez:2006uu}
\begin{align}
\mathcal{L}^{(int)} = \frac{1}{2}(\kappa_1  T^{\mu\nu}+\kappa_2 T\eta^{\mu\nu})h_{\mu\nu}  \label{inter}
\end{align}
when $\partial_\mu T^{\mu\nu}=0$. In particular, the coupling is Diff invariant for $\kappa_2=0$. This implies that the additional scalar mode mediates a  new gravitational  interaction with effective coupling
\begin{align}
\kappa_{eff}^2=-\frac{1}{\Delta b}\left(\kappa_2+\frac{1-a}{2}\kappa_1\right)^2    
\end{align}
with $\kappa_{eff}=8\pi G_{eff}$. However, in the $a_4=0$,  $a_5\neq 0$ case we have $a=1$ and provided the coupling to matter is Diff invariant i.e. 
$\kappa_2=0$, we get $\kappa_{eff}=0$, and the scalar mode is decoupled. 

Notice that the $f_\Lambda(g)$ term in \eqref{SG} plays the role of a potential term for the scalar mode $h$ which could provide a mass term. A priori, this term could be generated by radiative corrections even if it is not present at tree level. However, the shift symmetry of \eqref{modelterms} will protect against the generation of such terms so that we will restrict our analysis to the $f_\Lambda(g)=0$
case.

According to the above discussion, in this work we will concentrate on the
Diff invariant breaking induced by the $\mathcal{L}_5$ term. Notice that this term can 
be written as 
\begin{equation}
    \mathcal{L}_5 = -\frac{1}{4}g^{\mu \nu}(\partial_\mu \ln g)( \partial_\nu \ln g) 
\end{equation}
so that we can write the (non-linear) model under consideration as
\begin{align}
S_G=-\frac{1}{16\pi G}\int d^4x \left(f(g)R+ f_5(g)  \mathcal{L}_5\right)
\label{SG5}
\end{align}
Notice that this model does not deviate from GR at the linear level, and although it propagates an additional scalar graviton it is decoupled from matter if the matter coupling is Diff invariant. In addition for $a_5>0$ the scalar graviton is not a ghost. Beyond the Newtonian approximation, the breaking of Diff invariance induces deviations in the post-Newtonian parameters \cite{Alvarez:2009ga}. 
However, it can be seen that if  the integration measure of 
the Einstein-Hilbert term takes the Diff invariant expression i.e. $f(g)=\sqrt{g}$. 
then we recover the standard PPN parameters of GR, i.e.
\begin{align}
    \gamma_{PPN}=\beta_{PPN}=1
\end{align}
for arbitrary $f_5(g)$ \cite{Damour:1992we,Alvarez:2009ga}. 

Regarding the form of $f_5(g)$, for simplicity in the following, we will work with 
\begin{align}
    f_5(g)=a_5 \sqrt{g}
\end{align}
with constant $a_5>0$  corresponding to the global $GL(4,\mathbb{R})$ symmetry mentioned before. Notice that this symmetry protects the form of this term  against radiative corrections. 

Thus, putting all the above results together, a viable TDiff invariant gravitational model, 
which propagates an extra scalar graviton mode, decoupled from the conserved sources, is described by the total action
\begin{align}
S&=-\frac{1}{16\pi G}\int d^4x \sqrt{g}\left(\; R-  \frac{a_5}{4} g^{\mu \nu}(\partial_\mu \ln{g}) (\partial_\nu \ln{g})\right)\nonumber \\
&+\int d^4x \sqrt{g} \;\mathcal{L}_{m}
\label{total}
\end{align}
where $\mathcal{L}_{m}$ is the Diff invariant matter Lagrangian\footnote{Notice that for a Diff invariant matter sector, we do not expect radiative corrections from matter loops to the $\kappa_2$ coefficient
of the interaction Lagrangian \eqref{inter}.}. This model 
agrees with the unimodular bimode gravity discussed in \cite{Pirogov:2014lda}. 

Thus, interestingly, the model in \eqref{total} provides a description of the gravitational interaction that would be stable and 
indistinguishable from GR at the PPN level. Even though the theory behaves as GR in local 
gravity experiments, its non-linear dynamics can be very different. In particular, its cosmological evolution can differ from standard $\Lambda$CDM cosmology. It is precisely the aim of this work to analyze
the cosmological implications of this model.

\section{Modified Einstein equations}

Varying the total action in \eqref{total} with respect to the metric tensor we obtain
the corresponding Einstein equations
\begin{equation}
    G_{\mu \nu}  + a_5 \mathcal{M}_{\mu \nu}  = 8\pi G T_{\mu \nu}
\end{equation}
where

\begin{align}
\begin{split}
    \mathcal{M}_{\mu \nu} &= -\frac{1}{8}(\partial_\alpha\ln{g})(\partial_\beta\ln{g})(g_{\mu \nu}g^{\alpha \beta} + 2\delta^\alpha_\mu \delta^\beta_\nu) \\&- \frac{1}{2}g_{\mu \nu}\partial_\alpha (g^{\alpha \beta}\partial_\beta \ln{g})
\end{split}
\end{align}

Notice that because of the Diff invariance breaking, a priori, $\nabla_\mu \mathcal{M}^{\mu \nu} \neq 0$. However, since the matter sector is still Diff invariant and the energy-momentum tensor is conserved  $\nabla_\mu T^{\mu \nu} = 0$, 
we will have 
\begin{align}
\nabla_\mu \mathcal{M}^{\mu \nu} = 0  \label{cons}
\end{align}
on solutions of the Einstein equations.

\section{Modified Friedmann equations}

Let us now apply the above equations to cosmological backgrounds. Since  it is not possible
in general to fix coordinates in which $g_{00}=1$ with a TDiff transformation, we have to consider
a general form of the spatially homogeneous and isotropic Robertson-Walker metric \cite{Alvarez:2007nn}. We will work with flat spatial sections for simplicity
\begin{align}
ds^2=b^2(\tau)d\tau^2-a^2(\tau)d\vec x^2 \label{RW}
\end{align}
where now both $a(\tau)$ and $b(\tau)$ have to be obtained from  the  Einstein equations. 

The energy-momentum tensor for a perfect fluid reads
\begin{align}
\begin{split}
    &T_{\mu \nu} = (\rho + p)u_\mu u_\nu - pg_{\mu \nu}
\end{split}
\end{align}
and the energy conservation reads
\begin{equation}
\label{conservacion-e}
\rho' + 3\frac{a'}{a}(\rho + p) = 0
\end{equation}
where prime denotes derivation with respect to the coordinate time $d/d\tau$. 
On the other hand the Friedmann equation reads
\begin{equation}
    \left(\frac{a'}{ab} \right)^2 + \frac{a_5}{3} \mathcal{M}^0_{\;0} = \frac{8\pi G}{3}\rho
\end{equation}
where $\rho=\rho_M+\rho_R+\rho_\Lambda$ correspond to the total energy density. 
On the other hand the acceleration equation reads
\begin{equation}
\begin{split}
    \frac{a''}{ab^2} - \frac{a'b'}{ab^3} - \frac{a_5}{6}\left(\mathcal{M}^0_{\;0}-\mathcal{M}^i_{\;i}\right) = -\frac{4\pi G}{3}(\rho + 3p) \\ 
\end{split}
\end{equation}
where summation in $i$ is implicit. 

Changing to cosmological time $dt = b(\tau)d\tau$, the conservation equation 
\eqref{conservacion-e} takes the usual form
\begin{align}
  \dot \rho +3\frac{\dot a}{a}(\rho+p)=0  \label{consrho}
\end{align}
and the Friedmann equation reads
\begin{equation}
    \left(\frac{\dot{a}}{a} \right)^2 +  \frac{a_5}{3} \mathcal{M}^0_{\;0} = \frac{8\pi G}{3}\rho
    \label{Friedmann}
\end{equation}
whereas the acceleration equation takes the form
\begin{equation}
\begin{split}
    \frac{\ddot a}{a}  - \frac{a_5}{6}\left(\mathcal{M}^0_{\;0}+3\mathcal{M}\right) = -\frac{4\pi G}{3}(\rho + 3p) \\ 
\end{split}
\end{equation}
with 
\begin{equation}
\label{M500}
    \mathcal{M}^0_{\;0} = -3\left(\frac{\Ddot{a}}{a} + \frac{7}{2}\left[\frac{\dot{a}}{a} \right]^2 + \frac{1}{3}\frac{\Ddot{b}}{b} - \frac{1}{6}\left[\frac{\dot{b}}{b} \right]^2 + 2\frac{\dot{a}\dot{b}}{ab}  \right) 
\end{equation}
\begin{equation}
\label{M5xx}
    \mathcal{M}^{i}_{\;j} = -\mathcal{M}\delta^i_{\;j}
\end{equation}
where
\begin{equation}
     \mathcal{M}=3\left(\frac{\Ddot{a}}{a} + \frac{1}{2}\left[\frac{\dot{a}}{a} \right]^2 + \frac{1}{3}\frac{\Ddot{b}}{b} - \frac{1}{2}\left[\frac{\dot{b}}{b} \right]^2\right)
\end{equation}
Thus, we can define an effective energy density associated to the extra term as
\begin{align}
\rho_S=-\frac{a_5}{8\pi G}\mathcal{M}^0_{\;0}
\end{align}
and the corresponding effective pressure as 
\begin{align}
p_S=-\frac{a_5}{8\pi G}\mathcal{M}
\end{align}
which according to the conservation equation \eqref{cons} satisfy
\begin{align}
  \dot \rho_S +3H(\rho_S+p_S)=0  \label{scons}
\end{align}
The Hubble parameter takes the usual expression
\begin{align}
H= \frac{\dot{a}}{a}
\end{align}
whereas now we can define an additional Hubble parameter for the time
component
\begin{align}
H_b= \frac{\dot{b}}{b}
\end{align}
In terms of the new variables, the Friedmann and pressure equations read
\begin{align}
\label{hpunto}
        \dot{H} &= -\frac{a_5}{2}\left(H_b + 3H \right)^2 - 4\pi G(\rho + p)\\
        \dot{H}_b &= \frac{3 a_5}{2}\left(H_b + 3H\right)^2 - \frac{1}{2}\left(H_b + 3H \right)(H_b + 9H) \nonumber \\ &+ \frac{3}{a_5}H^2 + 4\pi G\left(3(\rho + p) - \frac{2}{a_5}\rho \right) \label{hbpunto}
\end{align}

Given the fact that ${\cal L}_5$ only depends on $g$, these equations can be written in an even simpler way by introducing the new variable
\begin{equation}
    H_g \equiv \frac{\dot{g}}{g}  = 2H_b + 6H
\end{equation}
so that we find
\begin{align}
    \dot{H}&= -\frac{a_5}{8}H_g^2 - 4\pi G(\rho + p)\label{Hdot} \\
    \dot{H}_g &= -\frac{1}{4}H_g(H_g+12H) + \frac{6}{a_5}\left( H^2-\frac{8\pi G}{3} \rho  \right) \label{Hbdot}
\end{align}
In terms of the new variables, the effective equation of state of the scalar mode can be written as 
\begin{equation}
    \omega_S =\frac{p_S}{\rho_S} = - 1 + \frac{a_5}{12}\frac{H_g^2}{\frac{8\pi G}{3}\rho_S}= - 1 + \frac{a_5}{12}\frac{H_g^2}{H^2-\frac{8\pi G}{3}\rho} \label{ws}
\end{equation}
where in the last step we have used \eqref{Hbdot} written in the Friedmann form
\begin{equation}
  H^2= \frac{8\pi G}{3} (\rho +  \rho_S) \label{FS}
\end{equation}
with
\begin{align}
\frac{8\pi G}{3}\rho_S=\frac{a_5}{6}\left(\dot{H}_g+\frac{1}{4}H_g(H_g+12H)\right)
\end{align}
Thus we see that if $H_g(t)=0$ then $\rho_S(t)=0$ and the extra scalar mode is not excited. On the other hand, from \eqref{ws} we see that for $\rho_S > 0$,  the condition $a_5>0$ implies $\omega_S \geq -1$, whereas $a_5 < 0$, which corresponds to a  scalar ghosts, implies a phantom effective equation of state $\omega_S \leq -1$. Notice also that the effective fluid cannot behave as
an exact cosmological constant, since that would imply $H_g(t)=0$ i.e.  $\rho_S(t)=0$.

Notice that in GR the usual Friedmann equation allows to solve
for the scale factor with a first order equation and the free parameters 
of the model are $(H_0, \Omega_M,\Omega_R)$, where using the cosmic sum rule for flat spatial sections $\Omega_\Lambda=1-\Omega_M-\Omega_R$. However,   
now we have a system of two second order equations and we
need an additional parameter
\begin{align}
    H_g^0=H_g(t_0)
\end{align}
in order to specify the cosmological model \footnote{Note that we can set $g(t_0)=1$ without loss of generality.}. Unlike the $H_0$ parameter which can be measured independently from the rest of cosmological parameters from low-redshift Hubble diagrams, this is not the case of $H_g^0$. However, it is always possible to measure it from the joint fit analysis with the rest of cosmological parameters with distance indicators from SNIa, BAO or CMB data. Moreover, no sum rule applies to the ordinary density parameters in this case since now there is an additional contribution $\rho_S$ in \eqref{FS}, so that the set of independent  cosmological parameters would be $(H_0, H_g^0, \Omega_M,\Omega_R,\Omega_\Lambda)$.

Finally, note that if initially the extra gravitational mode is not excited, i.e $\rho_S=0$, the cosmological evolution will be the same 
as in standard GR, i.e. very much as in the linearized regime, ordinary matter is not a source of the extra gravitational mode which remain decoupled from matter.

\section{Solutions with constant $\omega_S$}

In order to obtain explicit solutions, let us rewrite our system of equations \eqref{Hdot} and \eqref{Hbdot} as
\begin{align}
H^2&= \frac{8\pi G}{3} (\rho +  \rho_S) \label{H2}\\
H_g^2&= \frac{32\pi G}{a_5}(1+\omega_S)\rho_S \label{Hg2}
\end{align}
where in the last equation we have used conservation equations
\eqref{consrho} and \eqref{scons}. It is then straightforward to 
look for solutions with constant equation of state for the 
scalar fluid $\omega_S$. Thus taking the derivative of the second
equation and using \eqref{Hbdot} we find
\begin{align}
\rho_S(1-\omega_S)(H_g-6H(1+\omega_S))=0 \label{expl}
\end{align}
From these equations we can readily find explicit solutions

\begin{itemize}
\item {\bf $\Lambda$CDM solution}.  For any value of $a_5$, this solution corresponds to 
\begin{align}
    H_g &= 0 \label{LCDM1}\\
    H^2 &=\frac{8\pi G}{3}\rho \label{LCDM2}
\end{align}
so that $\rho_S$ vanishes and the scalar mode has no effect at all thus recovering the standard $\Lambda$CDM cosmology. In this case, the metric determinant is just a constant $g=$ const. Notice that the existence of this solution guarantees that, at the background level, the model can fit current observations of CMB, SNIa and BAO for any value of $a_5$ with at least the same accuracy as $\Lambda$CDM.
\item {\bf Stiff fluid solution ($\omega_S = 1$)}. For all $a_5>0$, this solution corresponds to 
\begin{align}
    H_g &=  \pm\sqrt{\frac{24}{a_5}\frac{8\pi G}{3}\rho_S} \label{stiff}\\
    H^2 &=\frac{8\pi G}{3}(\rho+\rho_S)
\end{align}
with $\rho_S\propto a^{-6}$, i.e.  the behaviour of the scalar mode is that of a stiff fluid.  
\item \textbf{Tracker solution}. If the matter sector only contains a single fluid with constant equation of state $\omega\neq -1$, a solution is present in which the scalar fluid 
tracks the matter behaviour with the same equation of state, 
i.e. 
\begin{align}
    \omega_S=\omega
\end{align}
so that from \eqref{expl}
\begin{align}
    H_g=6H(1+\omega) \label{tracker}
\end{align}
Substituing back in \eqref{Hg2} and using \eqref{H2} we obtain
the constant ratio between the two fluids
\begin{align}
\rho=\left(-1+\frac{1}{3a_5(1+\omega)}\right)\rho_S   \label{ratio}  
\end{align}
Note that $\left\vert\frac{\rho_S}{\rho}\right\vert\in [0,\infty)$, so that the effective fluid can be dominant over the tracked fluid.
Note also that as mentioned before, the $\omega=-1$ would imply $\rho_S=0$. In order to have $\rho_S > 0$ for $\rho > 0$  in \eqref{ratio}, we should have (for $a_5>0$)
        \begin{equation}
        a_5<\frac{1}{3(1+w)} 
        \end{equation}

\item {\bf Vacuum solution}.  This is a limiting case of the tracker solution when considering $\rho = 0$, and the only solution with constant $\omega_S$ in vacuum, apart from the general stiff solution \eqref{stiff}. Thus from \eqref{ratio} we get
\begin{align}
\omega_S^{\infty} = -1+\frac{1}{3a_5}    \label{winfinity}    
\end{align}
and
\begin{align}
H_g &= \frac{2}{a_5}H \label{vacratio}\\
H^2 &= H_0^2 \left(\frac{a}{a_0}\right)^{-\frac{1}{a_5}}
\end{align}
Notice that for  $a_5 = \frac{1}{6}$, this solution coincides with the positive branch of the stiff solution \eqref{stiff}. On the other hand for large $a_5$, the equation of state corresponds to a dark energy fluid. We will show that this solution corresponds to the asymptotic future limit of a quite general set of solutions, thus an asympototic dark energy behaviour generically requires large values of $a_5$.

\end{itemize}
\section{Effective equation of state}

In order to understand the phenomenology of the new term, it is useful to write down a differential equation system for $\omega_S$ and the  Hubble parameter with the scale factor $a$ as independent variable. Thus, using \eqref{ws} in \eqref{Hdot} and \eqref{Hbdot} we find for 
$a_5>0$ 
\begin{align}
    \frac{dH}{da} &= -\frac{3}{2a}H(\omega_S + 1) + \frac{4\pi G}{Ha}(\rho \,\omega_S - p)\\
    \frac{d \omega_S}{da} &= \frac{\omega_S - 1}{a}\left(3(\omega_S + 1) - K_{eff}\sqrt{\omega_S + 1}\right) \label{wseq}
\end{align}
where
\begin{equation}
    K_{\text{eff}}(a) = K\, \Omega_S^{1/2}(a)\quad ,  \quad K = \text{sgn}(H_g) \sqrt{\frac{3}{a_5}}
\end{equation}
with
\begin{align}
    \Omega_S(a)=1-\Omega_{tot}(a), \quad \Omega_{tot}(a)=\frac{8\pi G \rho(a)}{3H^2(a)}
\end{align}
Here $\Omega_S(a)$ is the relative abundance of $\rho_S$ at a given time. $K$ has two possible signs because for a given $\omega_S$ we have two possible signs for $H_g$, since $\omega_S+1 \propto H_g^2$ in \eqref{ws}. Thus the sign of $H_g$ divides the solutions in two branches: the branch with $K>0$ which has growing $g$, i.e. $H_g > 0$   and that with $K<0$ and contracting determinant $H_g < 0$. The point $H_g = 0$ which corresponds to $\omega_S = -1$ connects the two branches. Thus a complete solution is obtained by matching the two branches at the critical point. 

From \eqref{wseq} it is straightforward to find the stiff matter solution with constant $\omega_S(a) = 1$
mentioned before. Also, the previously mentioned vacuum solution would correspond to $\Omega_S(a)=1$,
so that  
\begin{align}
3(\omega_S + 1) - K\sqrt{\omega_S + 1}=0
\end{align}
whose real solution corresponds to a constant solution with $\omega_S(a) = \omega_S^{\infty} = -1+\frac{1}{3a_5}$. 

Equation \eqref{wseq} suggests that an additional solution with constant equation of state $w_S(a)=-1$ would exist. However, as mentioned before, that 
solution would correspond to $H_g(a)=0$ and accordingly vanishing effective
energy density.

In terms of  the density parameter for the extra contribution today $\Omega_S$, it is possible to write the effective equation of state parameter today $\omega_S^0=\omega_S(a=1)$ from \eqref{ws} as
\begin{align}
 \omega_S^0 = - 1 + \frac{a_5}{12}\frac{(H_g^0)^2}{H_0^2 \Omega_S}  
\end{align}
Thus, a dark energy behaviour today with $\omega_S^0<-1/3$ would require $a_5(H_g^0)^2<8H_0^2 \Omega_S$.

\section{Subdominant regime $\vert \rho_S\vert \ll \rho $ }
If the extra contribution is subdominant with respect to the ordinary energy components, we have $\vert\Omega_S(a)\vert\ll 1$ and  the equation for $\omega_S$ \eqref{wseq} reduces to
\begin{align}
    \frac{d \omega_S}{da} &= \frac{3(\omega_S^2 - 1)}{a}
\end{align}
whose solution reads
\begin{align}
w_S(a)=\frac{C-a^6}{C+a^6}
\end{align}
with $C\neq 0$ a real integration constant. From the conservation equation \eqref{scons}
we can write
\begin{align}
 \rho_S=\rho_S^0 \,e^{-3\int_1^a \frac{d\hat a}{\hat a}(1+w_S(\hat a))}   
\end{align}
so that 
\begin{align}
\rho_S=\frac{\rho_S^0}{(1+C)}\left(1+\frac{C}{a^6}\right)
\end{align}
i.e. the effective energy density is just the sum of a cosmological constant  and a stiff fluid contribution.
In addition, from \eqref{ws} we find
\begin{align}
H_g= \frac{{H_g^0}}{a^3}
\end{align}

\begin{figure}
\centering
    \includegraphics[width=\linewidth]{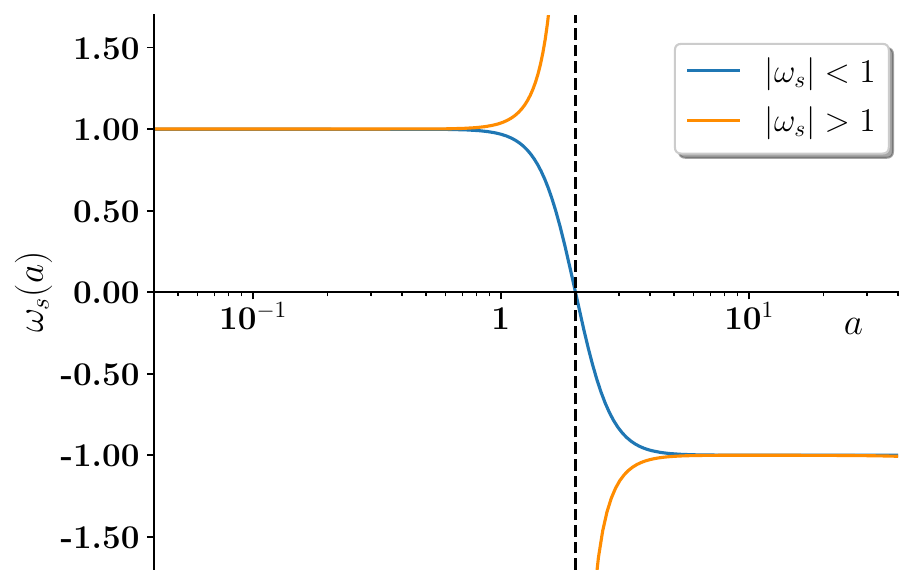}
    \caption{Evolution of the equation of state in the subdominant regime. The blue line corresponds to solutions with $\vert \omega_S\vert <1$ whereas the orange one corresponds to $\vert \omega_S\vert >1$. The vertical asymptote corresponds to the 
    change of sign of the effective energy density $\rho_S$ for the orange solution.}
    \label{fig:subdom}
\end{figure}

Thus depending on $C$, we have two different behaviours. As we can see in Fig. \ref{fig:subdom}, for $C>0$ and $\rho_S^0>0$, $\omega_S\in (-1,1)$ and the equation of state interpolates regularly from  an early stiff fluid behaviour $w_S=1$ and a late cosmological constant solution $w_S=-1$ with $\rho_S>0$. For $C<0$, we also have that the equation of state interpolates from  an early stiff fluid behaviour $w_S=1$ and a late cosmological constant solution $w_S=-1$ but with $w_S>1$ or $w_S<-1$ with $\rho_S$ changing sign at $a^6=-C$. Thus we see that the generic late time behaviour of the extra contribution is that of cosmological constant at least while the contribution is sub-dominant. Thus even a tiny Diff breaking generated by the extra term will freeze as a cosmological constant at late times.

\section{Dominant regime $ \rho_S \gg \rho$ }
In the dominant regime we can write the system as
\begin{align}
    \dot{H}&= -\frac{a_5}{8}H_g^2  \\
    \dot{H}_g &= -\frac{1}{4}H_g(H_g+12H) + \frac{6H^2}{a_5} 
\end{align}
which has a critical point at $H=H_g=0$ corresponding to Minkowski space-time and  three separatrixes.
\begin{align}
H_g&=\pm\sqrt{\frac{24}{a_5}}H, \;(w_S=1)\\
H_g&=\frac{2}{a_5}H, \; (w_S=w_S^\infty)
\end{align}
with $\omega_S^\infty$ given in \eqref{winfinity}. 
For $a_5=1/6$ the separatrix with $w_S=w_S^\infty$ coincides with 
one of the $w_S=1$ lines. Thus we have two possibilities 
$a_5\leq 1/6$ and $a_5>1/6$. 

\begin{figure}
\begin{center}
\includegraphics[scale = 0.4]{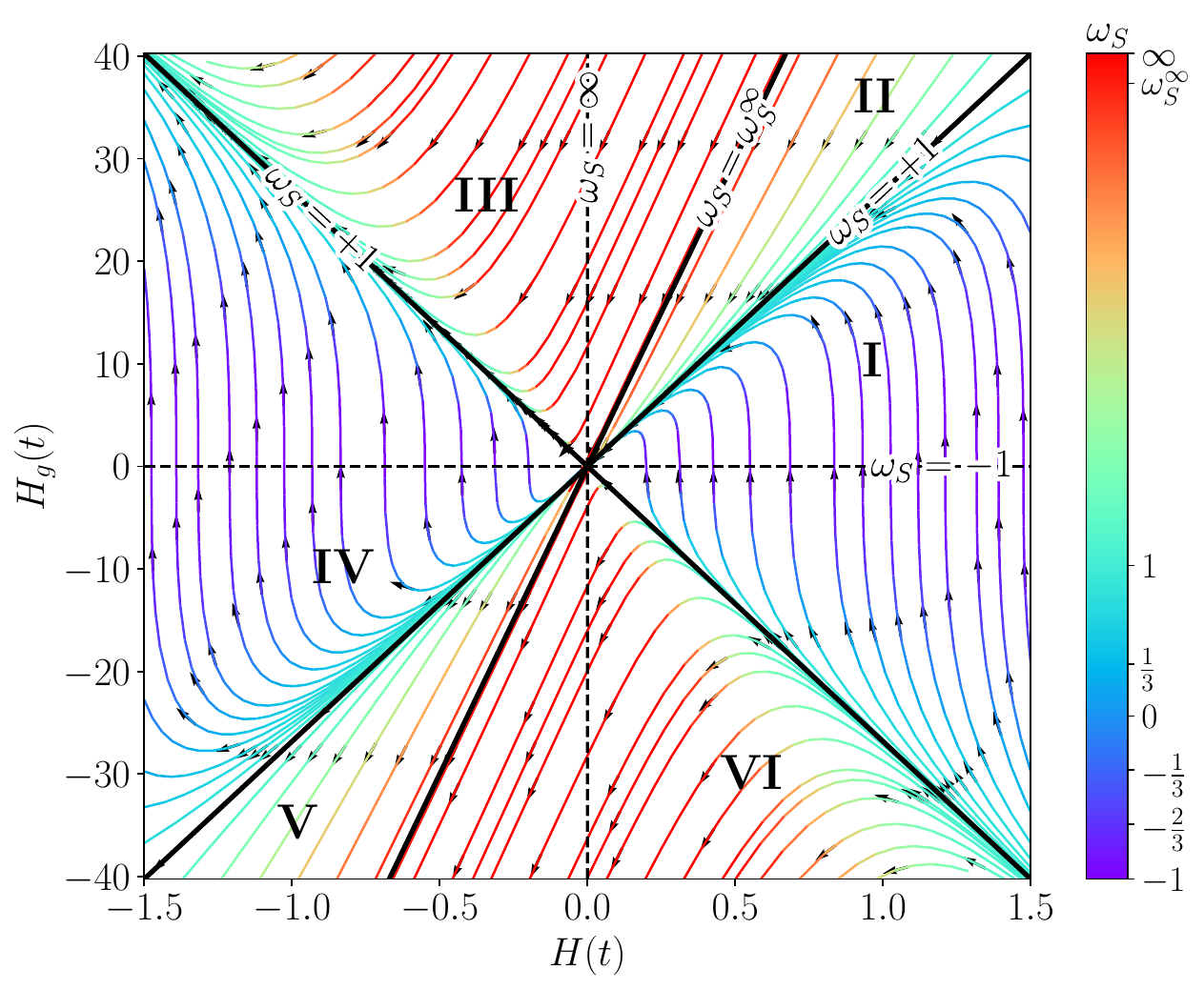}
\caption{Streamline plot in the dominant regime for $a_5<1/6$ ($a_5=1/30$) in $H_0$ units. The black lines correspond to the 
separatrixes $\omega_S=1$ and $\omega_S=\omega_S^\infty$ which delimit the six Regions I-VI. The dashed vertical and horizontal lines correspond to $\omega_S=\infty$ and $\omega_S=-1$ respectively. Curves with constant $\omega_S$ correspond to straight lines passing through the origin, $H_g=mH$ with $m=\pm\sqrt{(1+\omega_S)12/a_5}$. }
    \label{phpo1}   
\end{center}
\end{figure}

In Fig. \ref{phpo1} we plot the streamlines for $a_5<1/6$. We find six different regions delimited by the separatrixes:

\begin{itemize}
  
\item In Region I, we have expanding solutions which interpolate between $\omega_S=1$ in the remote past and $\omega_S=1$ in the asymptotic future, crossing the $\omega_S=-1$ line where $H_g$ changes from negative to positive sign. The Minkowskian critical point is reached in the  asymptotic future.

\item In Region II, we have expanding solutions with $H_g>0$, which interpolate between $\omega_S=\omega_S^\infty$ in the remote past and $\omega_S=1$ in the asymptotic future. The Minkowskian critical point is again reached in the  asymptotic future.

\item In Region III, we have  solutions with $H_g>0$, which interpolate between an expanding $\omega_S=\omega_S^\infty$ phase in the remote past and a contracting $\omega_S=1$ epoch in the asymptotic future which eventually recollapse. 

\item Region IV is the time reverse of Region I, in which we have contracting solutions which interpolate between $\omega_S=1$ in the remote past and $\omega_S=1$ in the asymptotic future, crossing the $\omega_S=-1$ line where $H_g$ changes from negative to positive sign and eventually recollapse. Solutions start out from the Minkowskian critical point asymptotically in the past. 

\item Region V is the time reverse of Region II.  We have contracting solutions with $H_g<0$, which interpolate between $\omega_S=1$ in the remote past and $\omega_S=\omega_S^\infty$  in the asymptotic future.  Solutions start out from the Minkowskian critical point and eventually recollapse.

\item Region VI is the time reverse of Region III. We have  solutions with $H_g<0$, which interpolate between an expanding $\omega_S=1$ in the remote past and a contracting $\omega_S=\omega_S^\infty$ phase  in the asymptotic future which eventually recollapse. 

\end{itemize}

\begin{figure}
\begin{center}
\includegraphics[scale = 0.4]{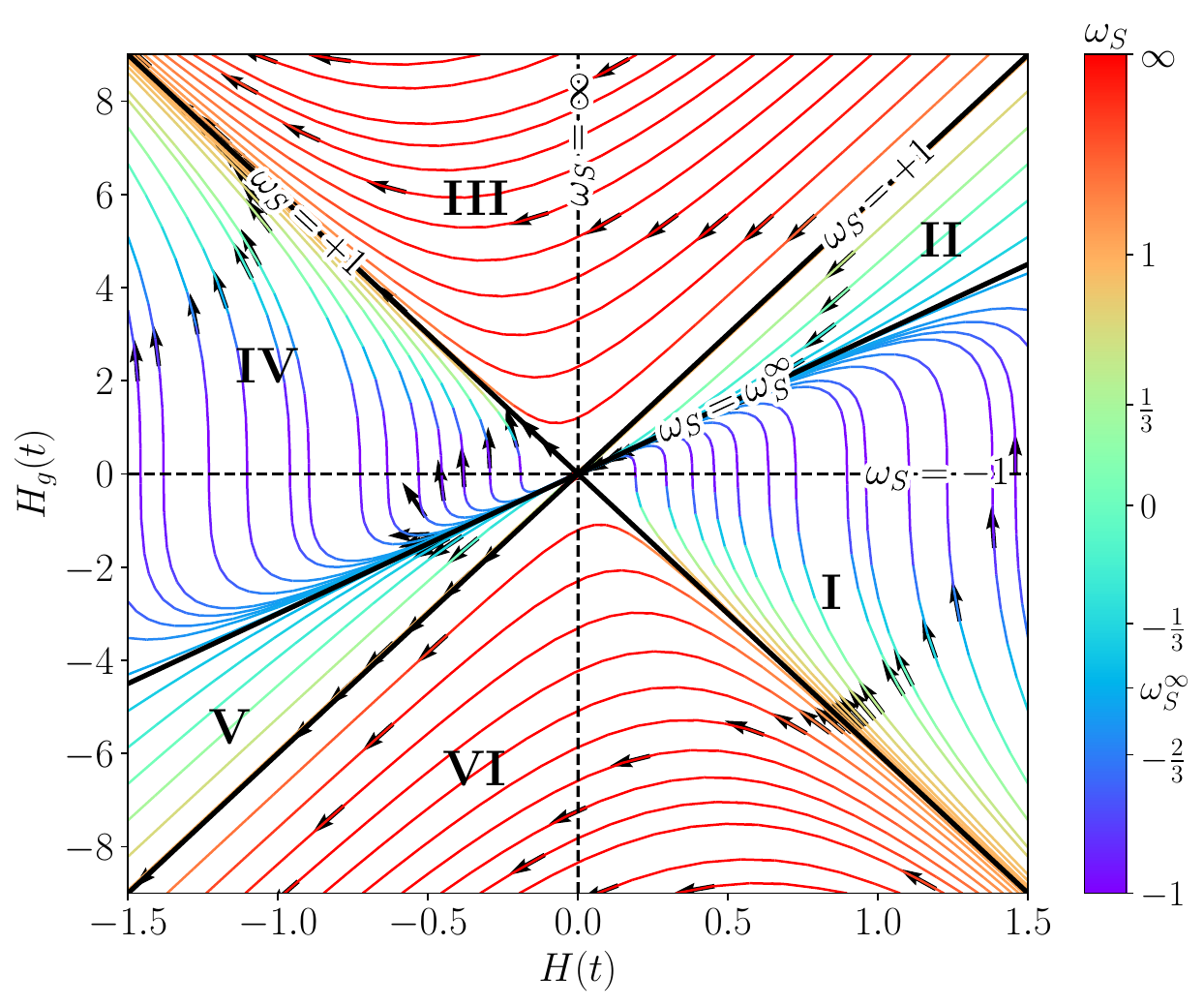}
 \caption{Streamline plot in the dominant regime for $a_5>1/6$ ($a_5=2/3$) in $H_0$ units. The black lines correspond to the 
separatrices $\omega_S=1$ and $\omega_S=\omega_S^\infty$. The dashed vertical and horizontal lines correspond to $\omega_S=\infty$ and $\omega_S=-1$ respectively. Curves with constant $\omega_S$ correspond to straight lines passing through the origin, $H_g=mH$ with $m=\pm\sqrt{(1+\omega_S)12/a_5}$. }
 \label{phpo2}  
\end{center}
\end{figure}

In Fig. \ref{phpo2} we plot the streamlines for $a_5>1/6$. We again find six different regions delimited by separatrixes: 
\begin{itemize}
\item In Region I, we have expanding solutions which interpolate between $\omega_S=1$ in the remote past and $\omega_S=\omega_S^\infty$ in the asymptotic future, crossing the $\omega_S=-1$ line where $H_g$ changes from negative to positive sign. The Minkowskian critical point is reached in the  asymptotic future.

\item In Region II, we have expanding solutions with $H_g>0$, which interpolate between $\omega_S=1$ in the remote past and $\omega_S=\omega_S^\infty$ in the asymptotic future. The Minkowskian critical point is again reached in the  asymptotic future.

\item In Region III, we have  solutions with $H_g>0$, which interpolate between an expanding $\omega_S=1$ phase in the remote past and a contracting $\omega_S=1$ epoch in the asymptotic future which eventually recollapse. 

\item Region IV is the time reverse of Region I, in which we have contracting solutions which interpolate between $\omega_S=\omega_S^\infty$ in the remote past and $\omega_S=1$ in the asymptotic future, crossing the $\omega_S=-1$ line where $H_g$ changes from negative to positive sign and eventually recollapse. Solutions start out from the Minkowskian critical point asymptotically in the past. 

\item Region V is the time reverse of Region II.  We have contracting solutions with $H_g<0$, which interpolate between $\omega_S=\omega_S^\infty$ in the remote past and $\omega_S=1$  in the asymptotic future.  Solutions start out from the Minkowskian critical point and eventually recollapse.

\item Region VI is the time reverse of Region III. We have  solutions with $H_g<0$, which interpolate between an expanding $\omega_S=1$ in the remote past and a contracting $\omega_S=1$ phase  in the asymptotic future which eventually recollapse. 

\end{itemize}
For $a_5=1/6$ regions II and V disappear and $\omega_S^\infty=1$.

Thus we see that unlike General Relativity in which the only solution in vacuum is Minkowski space-time (for flat spatial sections) in the broken Diff case, we have a wide range of cosmological solutions.

\subsection{Evolution of the equation of state}
In the dominant case with  $\Omega_{tot}(a)\ll 1$, we can take $K_{\text{eff}}=K$ so that
the equation for $\omega_S$ \eqref{wseq} reads
\begin{align}
    \frac{d \omega_S}{da} &= \frac{\omega_S - 1}{a}\left(3(\omega_S + 1) - K\sqrt{\omega_S + 1}\right) 
\end{align}
which can be explicitly integrated. Taking the variable $u = \sqrt{\omega_S + 1}$ gave us the following equation 

\begin{equation}
    \frac{du}{\frac{1}{2}(u^2-2)(3u-K)} = \frac{da}{a}
\end{equation}
The integrated expression has hyperbolic tangent functions as solutions, which explains the abrupt transitions in the evolution of the $\omega_S$ parameter.

Depending on the parameters values and initial conditions we can have different solutions which always evolve in two disconnected regions of $w_S$. 

\subsubsection{Solutions with $\vert  \omega_S\vert < 1$}

In this case the implicit solution reads for $a_5\neq 1/6$ i.e. $K^2\neq 18$
\begin{align}
\label{omegadea}
\begin{split}
    \ln{a} + C =  \frac{1}{K^2 - 18}\left(\sqrt{2}K \tanh^{-1}{\left[\sqrt{\frac{\omega_S+1}{2}}\right]}\right.& \\ + \left. 6\ln{\left| K-3\sqrt{\omega_S + 1} \right|} \vphantom{\frac{\sqrt{122}}{2}} - 3\ln{|\omega_S - 1|} \right)&
\end{split}
\end{align}
with $C$ an integration constant and $K = \pm \sqrt{\frac{3}{a_5}}$ correspond to  the two 
 branches of the solution. 

For $a_5 = 1/6$, corresponding to $K^2 = 18$, the solution reads
\begin{equation}
    \ln{a} + C = \pm\frac{1}{6}\left(\tanh^{-1}{\sqrt{\frac{\omega_S + 1}{2}}} + \frac{\sqrt{2}}{\sqrt{2}-\sqrt{\omega_S+1}}\right)
\end{equation}
which shows a similar behaviour as the $a_5<1/6$ case. These solutions corresponds to Regions I and IV for $a_5<1/6$ and Regions I, II, IV and V for $a_5>1/6$.

\begin{figure}
\centering
    \includegraphics[width=\linewidth]{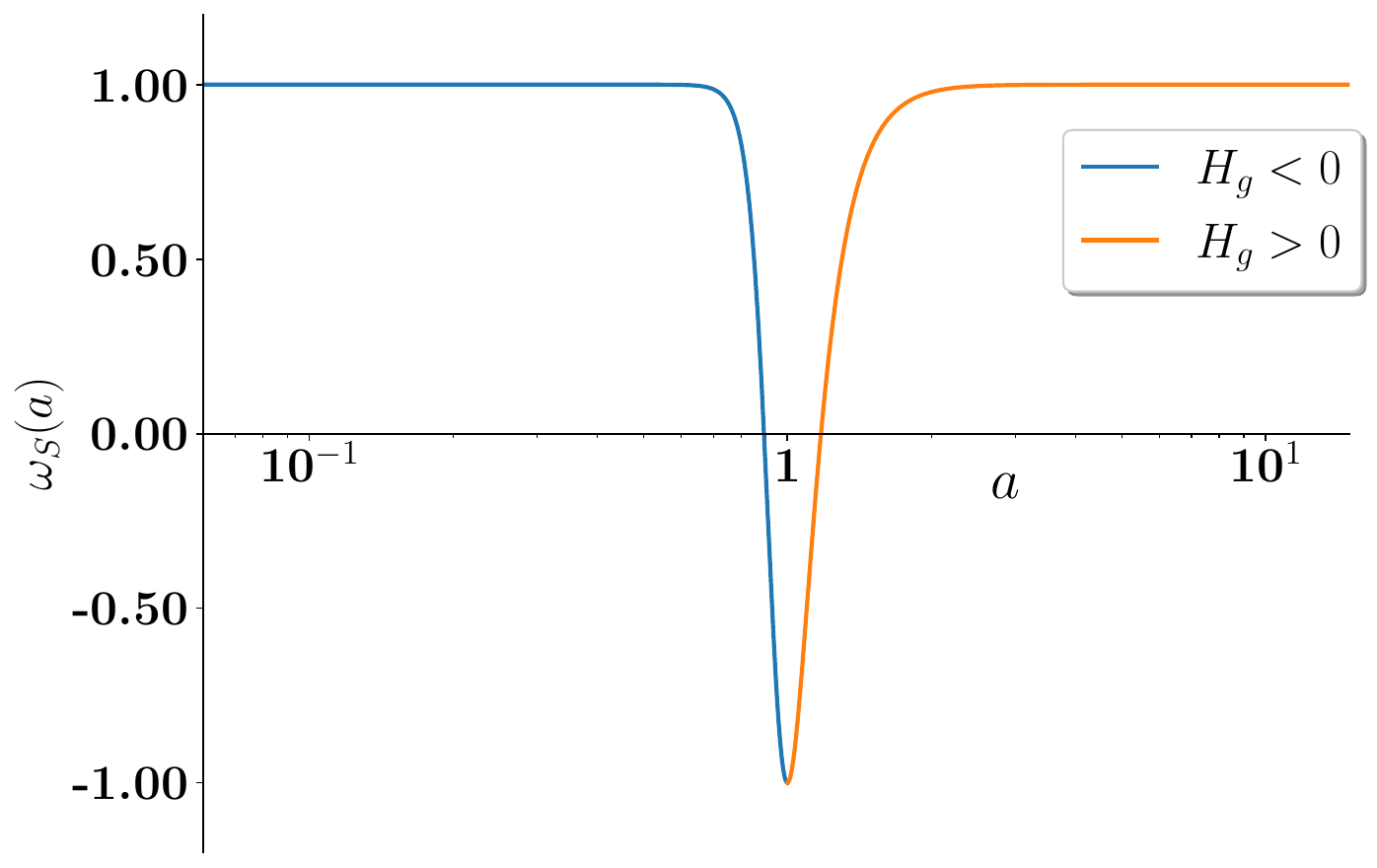}
    \caption{Evolution of the equation of state in the dominant regime for $a_5\leq 1/6$ in Region I. The blue line corresponds to the branch with $H_g<0$ whereas the orange one corresponds to $H_g >0$ branch.}
    \label{fig:domsmall}
\end{figure}

\begin{figure}
\centering
    \includegraphics[width=\linewidth]{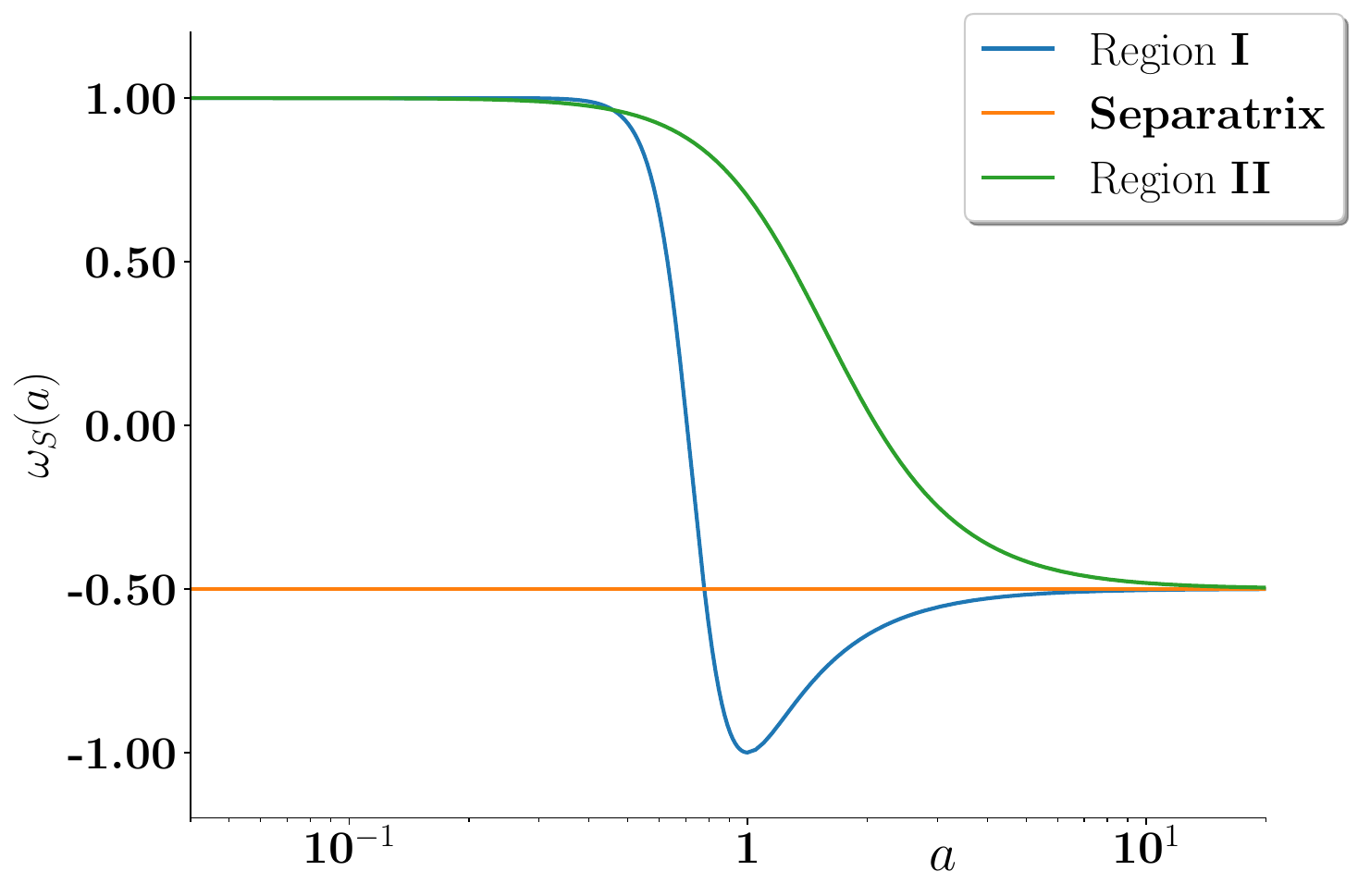}
    \caption{Evolution of the equation of state in the dominant regime for $a_5> 1/6$. In this particular case $a_5=2/3$ with $\omega_S^{\infty}=-1/2$. }
    \label{fig:domslarge}
\end{figure}

In Fig. \ref{fig:domsmall} we show the equation of state for $a_5=1/30$ corresponding to Region I in Fig. \ref{phpo1}, whereas in Fig. \ref{fig:domslarge} we show the equations of state for $a_5=2/3$ corresponding to Regions I, II and the separatrix with $\omega_S=\omega_S^{\infty}=-1/2$ of Fig. \ref{phpo2}.

    \subsubsection{Solutions with $\vert w_S\vert >1$}

The solution is extended to $\vert \omega_S \vert > 1$ by taking $\sqrt{\frac{2}{\omega_S + 1}}$ in the argument of the inverse hyperbolic tangent function in \eqref{omegadea}. Thus for $a_5\neq 1/6$ i.e. $K^2\neq 18$ we get 
\begin{align}
\label{omegadeawmas1}
\begin{split}
    \ln{a} + C =  \frac{1}{K^2 - 18}\left(\sqrt{2}K \tanh^{-1}{\left[\sqrt{\frac{2}{\omega_S + 1}}\right]}\right.& \\ + \left. 6\ln{\left| K-3\sqrt{\omega_S + 1} \right|} \vphantom{\frac{\sqrt{122}}{2}} - 3\ln{|\omega_S - 1|} \right)&
\end{split}
\end{align}
and for $K^2=18$
\begin{equation}
    \ln{a} + C = \pm\frac{1}{6}\left(\tanh^{-1}{\sqrt{\frac{2}{\omega_S + 1}}} + \frac{\sqrt{2}}{\sqrt{2}-\sqrt{\omega_S+1}}\right)
\end{equation}

\begin{figure}
\begin{center}
\includegraphics[scale = 0.35]{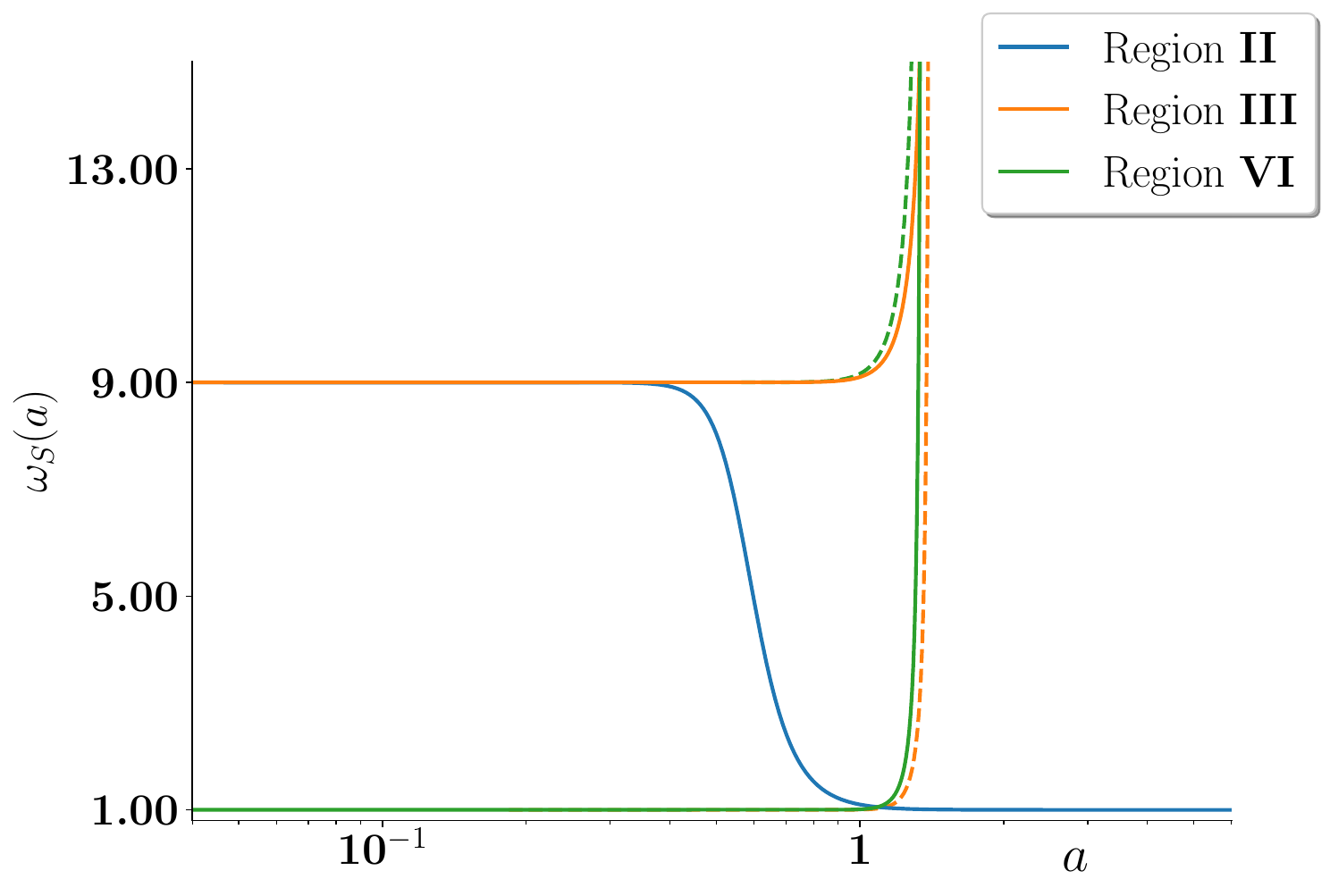}
\caption{Evolution of $\omega_S$ for $a_5<1/6$ ($a_5=1/30$) in contracting solutions (Regions II, III and VI). The full lines corresponds to the phase of the solution in which the universe is expanding (the evolution in time moves to growing $a$) whereas the dashed lines corresponds to the contracting phase and the evolution in time is thus towards decreasing $a$.}
    \label{dom_recollapse_1/30}   
\end{center}
\end{figure}

\begin{figure}
\begin{center}
\includegraphics[scale = 0.35]{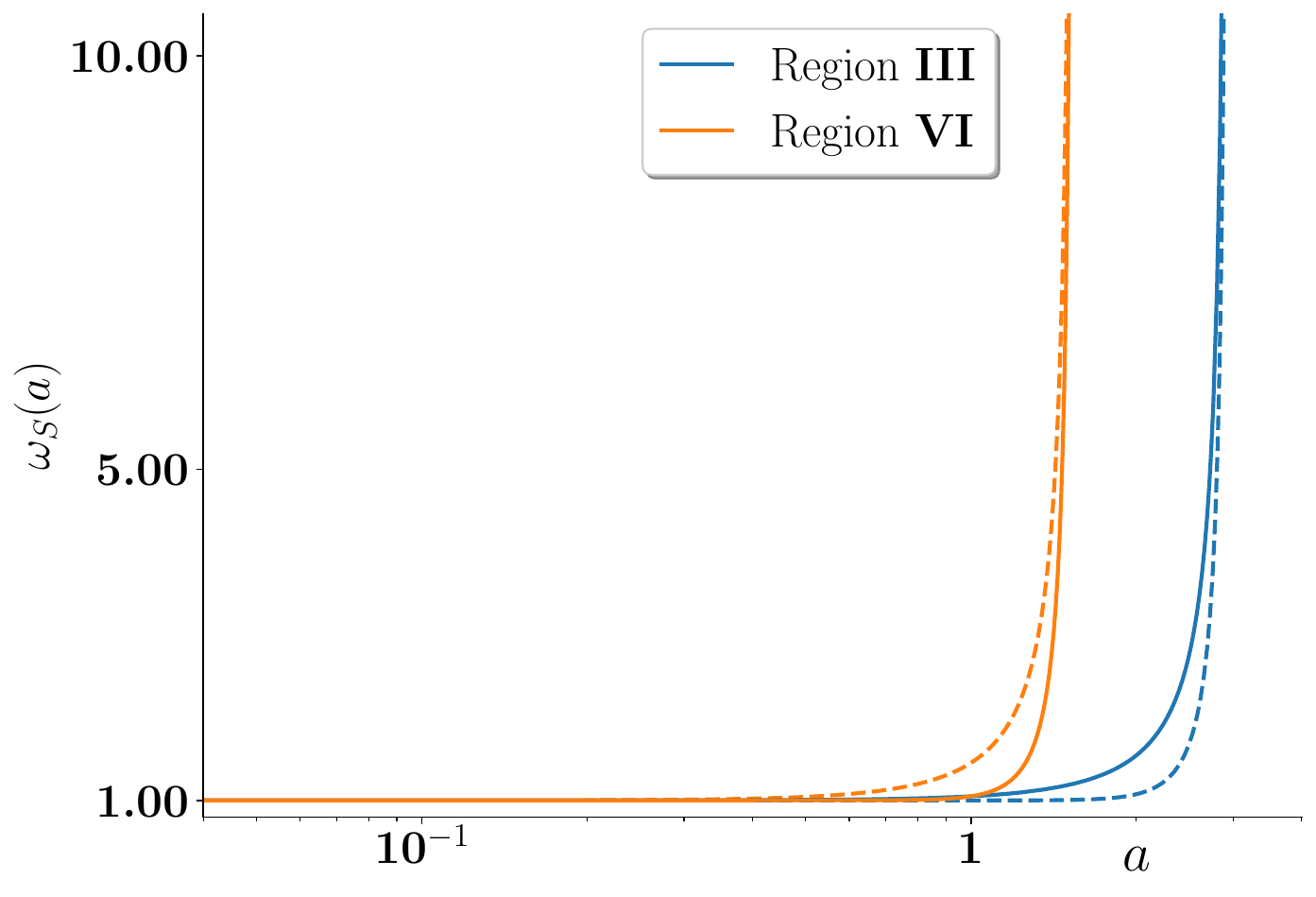}
\caption{Evolution of $\omega_S$ for $a_5>1/6$ ($a_5=2/3$) in contracting solutions (Regions III and VI). Full(dashed) lines as in  Fig. \ref{dom_recollapse_1/30}}
    \label{dom_recollapse_2/3}   
\end{center}
\end{figure}

In Fig. \ref{dom_recollapse_1/30} we show the equation of state for $a_5=1/30$ corresponding to Regions II, III and VI  of Fig. \ref{phpo1}, whereas in Fig. \ref{dom_recollapse_2/3} we show the equations of state for $a_5=2/3$ corresponding to Regions III and VI  of Fig. \ref{phpo2}.

\section{General cosmological evolution}
After exploring the limiting cases corresponding to dominant and subdominant scalar contribution with respect to the matter content, in this section we will consider the general case in which we cannot neglect any of the two contributions.

For a set of $n$ perfect fluids with constant equations of state $p_i=\omega_i\rho_i$, $i=1\dots n$, the equations for $H$, $H_g$ can be rewritten using conservation equation for the different matter components  as an autonomous system of $2+n$ dimensions
\begin{align}
        &\dot{H} = -\frac{a_5}{8}H_g^2 - \frac{3}{2}\sum\limits_{i = 1}^{n} H_i^2(\omega_i + 1) \label{s1}
        \\ 
        &\dot{H}_g = -\frac{1}{4}H_g(H_g + 12H) + \frac{6}{a_5}\left( H^2 - \sum\limits_{i=1}^{n} H_i^2 \right) \label{s2}
        \\
        &\dot{H}_i = -\frac{3}{2}H(\omega_i + 1)H_i, \;\; i=1\dots n  \label{s3}
\end{align}
where we have defined
\begin{align}
H_i^2 = \frac{8\pi G}{3}\rho_i, \;\; i=1\dots n 
\end{align}

As in the dominant case, the system has one critical point $H=H_g=H_i=0$,  $i=1\dots n$, which corresponds to the 
Minkowski solution. In addition, surfaces of constant $\omega_S$ are given by
\begin{align}
H^2-\sum_i H_i^2-\frac{a_5}{12(1+\omega_S)}H_g^2=0 
\end{align}
so that it can be seen that the only separatrix surface corresponds to $\omega_S=1$ and splits the space of solutions 
in two disconnected regions with  $|\omega_S| > 1$ and $|\omega_S| < 1$ respectively.

\begin{figure}
\begin{center}
\includegraphics[scale = 0.4]{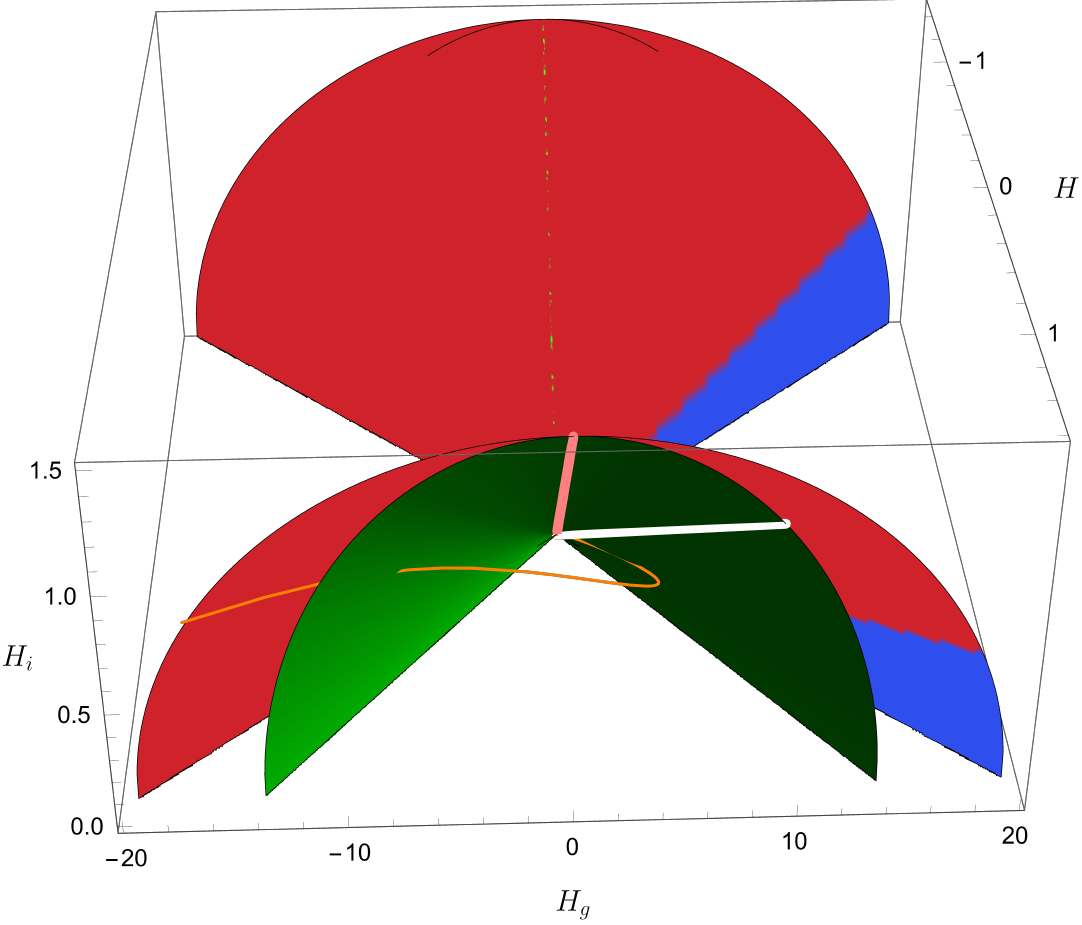}
\caption{Configuration space for the case with one fluid with equation of state $\omega_1=0$ and $\Omega_1=0.36$. The red/blue cone represents the separatrix i.e.  the surface with constant $\omega_S=1$. The two sheets of the cone correspond to expanding $H>0$ or contracting $H<0$ solutions. In the red zone the separatrix acts as a repulsor, whereas it is an attractor in the blue zone. The green cone corresponds to $\omega_S=0$. The pink straight line corresponds to the standard solution in GR for a matter dominated universe \eqref{LCDM1} and \eqref{LCDM2}
 which separates the $H_g>0$ from the $H_g<0$ regions, 
and the white line is the tracker solution \eqref{tracker}. The curve corresponds to a particular example solution which approaches the tracker asymptotically.}
    \label{general_sol}   
\end{center}
\end{figure}

Let us consider for simplicity the case with a single matter fluid i.e. $n=1$, with constant equation of state $\omega_1<1$. In this case, 
we will consider the two solutions regions: 
\begin{itemize}
    \item $|\omega_S| < 1$. In this case if
\begin{align}
    \omega_S^\infty=-1+\frac{1}{3a_5}>\omega_1
\end{align}
then we find that the extra component will approach asymptotically the tracker solution \eqref{tracker}. If $\omega_1\geq 1$, then the extra component will tend to the stiff fluid separatrix in the asymptotic future. 

In the opposite case in which 
\begin{align}
    \omega_S^\infty=-1+\frac{1}{3a_5}<\omega_1
\end{align}
then the solution tends to the dominant case with $H_i=0$ discussed in
Section VIII. As a matter of fact, the solutions in Figs. \ref{phpo1} and 
\ref{phpo2} correspond to the $H_i=0$ plane of Fig. \ref{general_sol}.

\item $|\omega_S| > 1$. In this case, generically the solution interpolates between an asymptotic stiff or $\omega_S^\infty$ fluid both  in the past and in the future eventually recollpasing. In the $a_5>1/6$
the interpolation is only between stiff fluid in the past and stiff fluid in the future.

\end{itemize}

\begin{figure}
\begin{center}
\includegraphics[scale = 0.35]{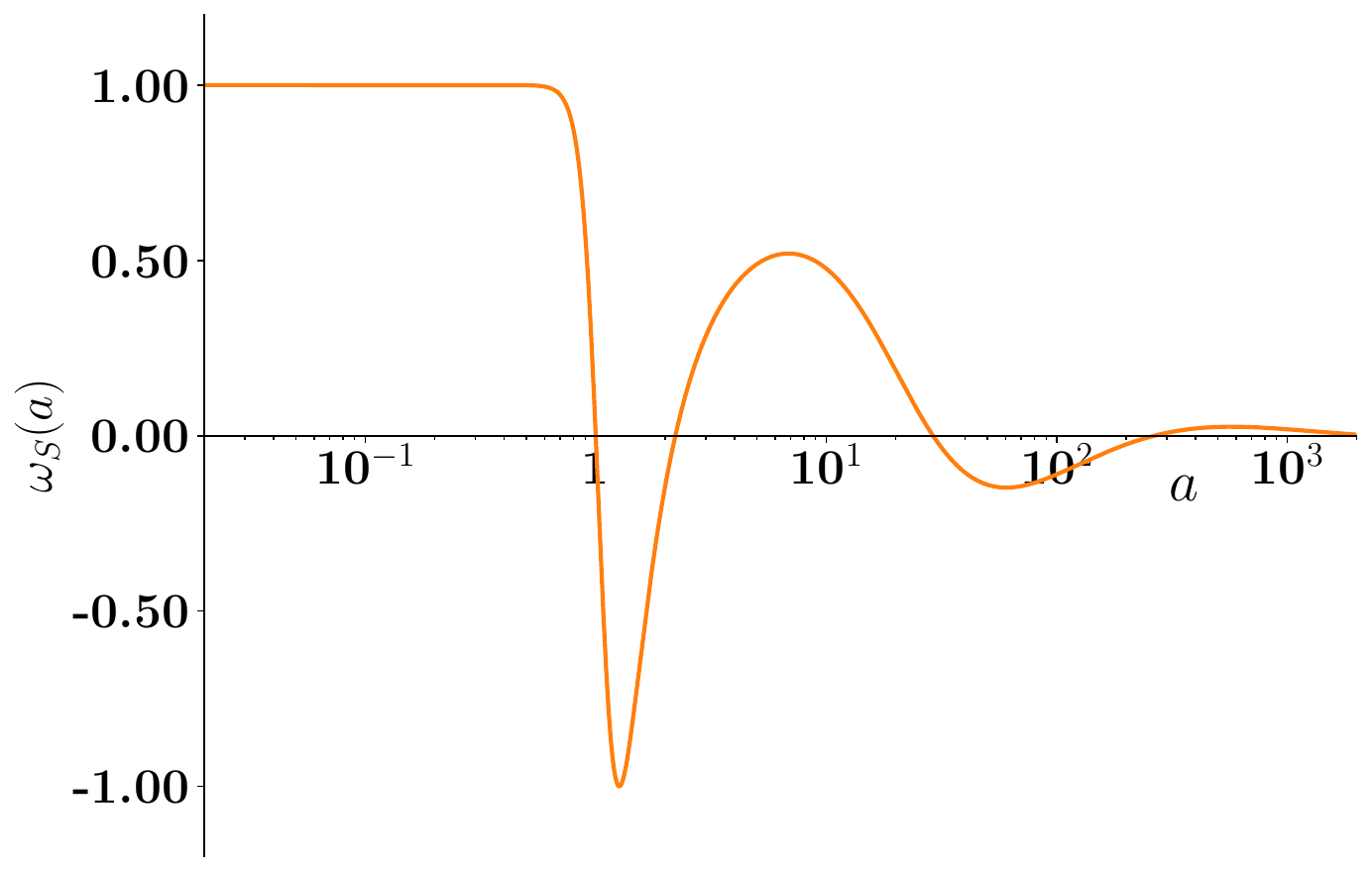}
\caption{Evolution of the effective equation of state $\omega_S$ in the case with one fluid with $\omega_1=0$ and $\Omega_1=0.36$ corresponding to the orange curve in Fig. \ref{general_sol}. We see 
how the solution tends to the tracker solution with $\omega_S=0$.}
    \label{sol_gen_ws}   
\end{center}
\end{figure}

In Fig. \ref{general_sol} we show as an example, the evolution 
of a solution in a simple case with $a_5=1/7$ in which we have only one matter component with equation of state $\omega_1=0$ corresponding to non-relativistic matter. The red/blue cone represents the separatrix i.e.  the surface with constant $\omega_S=1$. The two sheets of the cone correspond to expanding $H>0$ or contracting $H<0$ solutions. In the red region, the separatrix acts as a repulsor, whereas it is an attractor in the blue region. The green cone corresponds to $\omega_S=0$. The pink straight line corresponds to the standard solution in GR for a matter dominated universe \eqref{LCDM1} and \eqref{LCDM2}
 which separates the $H_g>0$ from the $H_g<0$ regions, 
and the white line is the tracker solution \eqref{tracker}. As we can 
see, the solution evolves from  $\omega_S=1$ in the past, crosses the
$\omega_S=0$ surface and approaches the tracker solution in the future. 
In Fig. \ref{sol_gen_ws}, we see the evolution of the effective equation 
of state for this particular solution.

\section{Stability of solutions}
In previous sections we have identified the existence of certain tracker solutions and other asymptotic behaviours. In this section we will prove that indeed they are stable attractors. With that purpose, let us consider small perturbations in  the general system of equations  \eqref{s1}, \eqref{s2} and \eqref{s3}. After linearizing the system we introduce the new variable $N=\ln a$, so that $\dot H=H H'$ where $'=d/dN$, and use the notation $\delta \hat H=\frac{\delta H}{H}$,
$\delta \hat H_g=\frac{\delta H_g}{H}$ and $\delta \hat H_i=\frac{\delta H_i}{H}$, so that we can write:

\begin{align}
    \delta \hat H'&=-\frac{H'}{H}\delta \hat H -\frac{a_5 H_g}{4H}\delta \hat H_g-3\sum_i (1+\omega_i) \frac{H_i}{H}\delta \hat H_i \\
    \delta \hat H_g'&=\left( -3\frac{H_g}{H}+\frac{12}{a_5}\right)\delta \hat H-\left(\frac{H_g}{2H}+\frac{H'}{H}+3\right)\delta \hat H_g\nonumber \\
    &-\frac{12}{a_5}\sum_i\frac{H_i}{H}\delta \hat H_i \\
    \delta \hat H_i'&=-\left(\frac{3}{2} (1+\omega_i)+\frac{H'}{H}\right)\delta \hat H_i
    -\frac{3}{2}(1+\omega_i)\frac{H_i}{H}\delta \hat H, \nonumber \\ \;\; i&=1\dots n
\end{align}

Let us then study the stability of some particular solutions.
\begin{itemize}
    \item {\bf Vacuum solution}. This corresponds to
    \begin{align}
    H_i&=0, \;\; i=1\dots n \\
        \frac{H_g}{H}&=\frac{2}{a_5}\\
        \frac{H'}{H}&=-\frac{1}{2a_5}
    \end{align}
    as shown in \eqref{vacratio}.

In this case, the system for the $\vec V= (\delta \hat H, \delta \hat H_g)$ variables decouple and can be written as  $\vec V'=M \vec V$ with
\begin{align}
    M= \begin{pmatrix}
        \frac{1}{2a_5} & -\frac{1}{2}\\  \frac{6}{a_5} & - \frac{1}{2a_5}-3 \end{pmatrix}
\end{align}

The corresponding eigenvalues are 
\begin{align}
    \lambda_1 &= -\frac{1}{2a_5}\\
    \lambda_2 &= -3 + \frac{1}{2a_5}
\end{align}
Thus we see that the vacuum solution  i.e. the solution with  dominant $\rho_S$ and 
$\omega_S=\omega_S^\infty$, is stable provided
 $a_5 > \frac{1}{6}$ so that $\lambda_{1,2}<0$, in agreement with the 
 streamline plots in Fig. \ref{phpo1}.

 The eigenvalues corresponding  to the $\delta \hat H_i$ variables are
\begin{align}
 \lambda_i = \frac{3}{2}(\omega_S^\infty - \omega_i),\;\; i=1\dots n 
\end{align}
 which indicate that stable solutions with $\lambda_i<0$ correspond to fluids
 with $\omega_i>\omega_S^\infty$ which ensure the background fluids to remain subdominant with respect to $\rho_S$. 
 \item {\bf Tracker solution}. This 
 corresponds to solutions in which there is  one fluid with constant equation of state $\omega_j$ which is tracked by the scalar fluid $(\omega_S=\omega_j)$, so that from \eqref{tracker} and \eqref{ratio} we have
 \begin{align}
  H_i&=0, \;\; i\neq j \\   
  \frac{H_j}{H}&=\sqrt{1-3a_5(1+\omega_j)}\\
 \frac{H_g}{H}&=6(1+\omega_j)\\
        \frac{H'}{H}&=-\frac{3}{2}(1+\omega_j)
 \end{align}
The system of equation corresponding to the variables $\vec V= (\delta \hat H, \delta \hat H_g, \delta \hat H_j)$ decouple from the rest so that the matrix can be written as 
\begin{equation}
        M_{\text{tracker}} = \begin{pmatrix}
            M_{\text{tracker}}^{3\times 3} & 0 \\ 0 & M_{\text{tracker}}^{(n-1)\times (n-1)}
        \end{pmatrix}
    \end{equation}
with
\begin{widetext}
    \begin{equation}
     M_{\text{tracker}}^{3\times 3} = \begin{pmatrix}
        \frac{3}{2}(\omega_j + 1) & -\frac{3}{2}a_5(\omega_j + 1) & -3(\omega_j + 1)\sqrt{1-3a_5(\omega_j + 1)} \\ -18(\omega_j + 1) +\frac{12}{a_5} & -\frac{3}{2}(\omega_j + 3) & -\frac{12}{a_5}\sqrt{1-3a_5(\omega_j + 1)} \\ -\frac{3}{2}(\omega_j + 1)\sqrt{1-3a_5(\omega_j + 1)} & 0 & 0
    \end{pmatrix}
\end{equation}
whereas the lower box is a diagonal matrix given by
\begin{equation}
     M_{\text{tracker}}^{(n-1)\times (n-1)} =\frac{3}{2} \begin{pmatrix}
        (\omega_j-\omega_1 ) & 0 & 0 & 0 & \dots &0 \\  
        \vdots & \vdots & \vdots & \vdots &\dots & \vdots \\
        0 &\dots  & (\omega_j -\omega_{j-1})& 0  &\dots &0 \\
        0 &  \dots &0 & (\omega_j-\omega_{j+1} ) & \dots & 0\\
        \vdots & \vdots & \vdots & \vdots &\dots & \vdots \\
        0 &  0 & 0 &0 & \dots & (\omega_j-\omega_{n}) 
    \end{pmatrix}
\end{equation}

The eigenvalues of this second matrix are trivial and stability $\lambda_{i\neq j} < 0$ imposes that the rest of background fluids are subdominant with respect to the tracked one.
    
 The three eigenvalues corresponding to the 
 $M_{\text{tracker}}^{3\times 3}$ matrix are
 \begin{align}
   \lambda_{1,2}&= -\frac{3}{4} \left(1-\omega_j\pm\sqrt{(1-\omega_j) \left(24 a_5
   (\omega_j+1)^2-9 \omega_j-7\right)}\right)\\
   \lambda_3&=-\frac{3}{2}(1+\omega_j)
\end{align}
Thus it can be seen that for $a_5\leq 1/6$
and $-1<\omega_j<1$ the three eigenvalues are negative so that the tracker solution is stable. On the other hand, for $a_5>1/6$, the stability condition is satisfied for $-1<\omega_j<\omega_S^\infty=-1+\frac{1}{3a_5}<1$
\end{widetext}

\end{itemize}

\section{Discussion and conclusions}

In this work we have considered the possibility of finding gravity models quadratic in metric derivatives which could break Diff invariance but are consistent with local gravity tests at the PPN level. We have identified such models as TDiff models which propagate an additional 
(non-ghost) massless scalar graviton mode, in addition to the standard, massless spin-2 mode and in which the matter sector is Diff invariant.
The Diff invariant coupling to matter ensures that the scalar graviton mode is
not sourced by matter fields and remains decoupled. 
The symmetries of the model protect the structure of the couplings from radiative corrections.  Notice that unlike other modifications of General Relativity, the breaking of Diff invariance allows to build local  gravity actions quadratic in metric derivatives without the introduction of additional gravitational fields.

Even though the model is indistinguishable from GR in the weak field approximation, its non-linear behaviour can be different. This suggests that cosmology is the perfect arena to test possible smoking guns of the model. However, the detailed analysis of the  full modified Friedmann equations shows that the 
extra terms associated to the new gravitational degree of freedom are 
not excited by the matter energy-momentum tensor. In other words, the 
new contributions could only have a primordial origin, for instance, from quantum fluctuations in the early universe. 
Thus, if the extra degree of freedom is initially not excited, the 
model recovers standard $\Lambda$CDM cosmology. However, once it is produced it can affect the cosmological evolution. Thus we have considered two main regimes: when the new contribution is negligible with respect to the ordinary background energy densities, we have shown that the 
effective energy density of the scalar mode behaves as a cosmological
constant. In other words, it freezes in the early universe and could have survived until present even for tiny primordial amplitudes, thus  providing a 
natural mechanism for dark energy generation. On the other hand,  when the new scalar contribution starts to dominate, we find two different behaviours depending on the relative size of the gravitational coupling constants of the new term compared to the Newton constant. This ratio  is in fact controlled by the $a_5$ parameter. Thus for
small $a_5$,  the effective energy density of the new contribution  $\rho_S(a)$ tracks that of the dominant background
fluid $\rho(a)$. This tracking behaviour could even be such that
$\rho_S(a)/\rho(a)>1$. On the opposite limit, for $a_5\gg 1$, the effective equation of state would be close to $\omega_S\simeq -1$ thus providing a 
natural dark energy candidate. Notice that it is  precisely this case with $a_5\gg 1$ the most interesting one from a phenomenological point of view, since the extra mode would behave as dark energy from the early universe. Indeed, assuming it was subdominant in the early universe, it would  have evolved as a cosmological constant $\omega_S\simeq -1$ during radiation and matter eras, and when it started to dominate it would have made a transition to  $\omega_S=-1+1/(3a_5)$. As shown before this evolution would be stable throughout the whole cosmic evolution.

Apart from these limits, we have shown that depending on the initial conditions and the value of $a_5$, other solutions are possible in which the universe evolves from an expanding to a  contracting phase, eventually recollapsing. 

\section{Prospects}
In this work we have limited ourselves to the analysis of the
homogeneous cosmological background. Primordial cosmological
perturbations of the new scalar degree of freedom could have some 
impact on structure formation and CMB anisotropies, thus providing alternative means to test the model. 
However, the breaking of Diff symmetry implies that the usual cosmological perturbation analysis in General Relativity cannot be straightforwardly applied. 
In particular, the usual gauge choices cannot be directly imposed (see \cite{Maroto:2023toq} for an introduction to perturbation theory in TDiff models) so that a direct implementation in Boltzmann codes such as CLASS or CAMB is not possible. The perturbations analysis will be presented in a forthcoming work. In addition, the
perturbed action would allow us to develop the quantization program for the extra scalar mode, which will allow to compute its primordial power spectrum generated during inflation. On the other hand, a confrontation of the model with current observations of SNIa, CMB and BAO will require to extend the usual 6 parameters likelihood analysis  of $\Lambda$CDM by including the three additional parameters $(a_5,H^0_g,\Omega_S)$ characterizing the TDiff model and will also be presented elsewhere. 

Apart from cosmology, the full non-linear behaviour of the model  could be tested in different contexts, for example in astrophysical scenarios with strong gravitational fields such as those associated to compact objects or black holes. 

\acknowledgements{A.G.BM acknowledges support from the Comunidad de Madrid INVESTIGO CT36/22-13-UCM-INV (NextGenerationEU). A.G.BM also acknowledges support from IPARCOS-UCM Master Thesis grants 2021-2022}. This work has been supported by the MICIN (Spain) project PID2019-107394GB-I00 (AEI/FEDER, UE).

\bibliographystyle{apsrev4-1}
\bibliography{bibliography}

\end{document}